\RequirePackage{lineno}
\documentclass[aps,prd,twocolumn,groupedaddress,superscriptaddress,nofootinbib]{revtex4-1}
\usepackage{graphicx}
\usepackage{amsmath}
\usepackage{float}
\usepackage{hyperref}
\usepackage{url}

\begin{document}


\title{Extending light WIMP searches to single scintillation photons in LUX}



\author{D.S.~Akerib} \affiliation{Case Western Reserve University, Department of Physics, 10900 Euclid Ave, Cleveland, OH 44106, USA} \affiliation{SLAC National Accelerator Laboratory, 2575 Sand Hill Road, Menlo Park, CA 94205, USA} \affiliation{Kavli Institute for Particle Astrophysics and Cosmology, Stanford University, 452 Lomita Mall, Stanford, CA 94309, USA}
\author{S.~Alsum} \affiliation{University of Wisconsin-Madison, Department of Physics, 1150 University Ave., Madison, WI 53706, USA}  
\author{H.M.~Ara\'{u}jo} \affiliation{Imperial College London, High Energy Physics, Blackett Laboratory, London SW7 2BZ, United Kingdom}  
\author{X.~Bai} \affiliation{South Dakota School of Mines and Technology, 501 East St Joseph St., Rapid City, SD 57701, USA}  
\author{J.~Balajthy} \affiliation{University of California Davis, Department of Physics, One Shields Ave., Davis, CA 95616, USA}  
\author{A.~Baxter} \affiliation{University of Liverpool, Department of Physics, Liverpool L69 7ZE, UK}  
\author{P.~Beltrame} \affiliation{SUPA, School of Physics and Astronomy, University of Edinburgh, Edinburgh EH9 3FD, United Kingdom}  
\author{E.P.~Bernard} \affiliation{University of California Berkeley, Department of Physics, Berkeley, CA 94720, USA}  
\author{A.~Bernstein} \affiliation{Lawrence Livermore National Laboratory, 7000 East Ave., Livermore, CA 94551, USA}  
\author{T.P.~Biesiadzinski$^*$}
\affiliation{Case Western Reserve University, Department of Physics, 10900 Euclid Ave, Cleveland, OH 44106, USA} \affiliation{SLAC National Accelerator Laboratory, 2575 Sand Hill Road, Menlo Park, CA 94205, USA} \affiliation{Kavli Institute for Particle Astrophysics and Cosmology, Stanford University, 452 Lomita Mall, Stanford, CA 94309, USA}
% ...

\collaboration{LUX collaboration}

\date{\today}


\begin{abstract}
We present a novel analysis technique for liquid xenon time projection chambers that allows for a lower threshold  by relying on events with a prompt scintillation signal consisting of single detected photons. The energy threshold of the LUX dark matter experiment is primarily determined by the smallest scintillation response detectable, which previously required a 2-fold coincidence signal in its photomultiplier arrays, enforced in data analysis. The technique presented here exploits the double photoelectron emission effect observed in some photomultiplier models at vacuum ultraviolet wavelengths. We demonstrate this analysis using an electron recoil calibration dataset and place new constraints on the spin-independent scattering cross section of weakly interacting massive particles (WIMPs) down to 2.5~GeV/c$^2$ WIMP mass using the 2013 LUX dataset. This new technique is promising to enhance light WIMP and astrophysical neutrino searches in next-generation liquid xenon experiments.
\end{abstract}


\maketitle


\modulolinenumbers[1]

\section{Introduction\label{sec:intro}}

Experiments searching for the scattering of weakly interacting massive particles (WIMPs), which are hypothesized to constitute the dark matter (DM) content of the universe, probe a variety of rare processes leading to $\mathcal{O}$(keV) energy transfers to ordinary matter. Specifically, several theories predict the existence of light DM particles (close to the proton mass) in addition to the standard thermally-produced WIMPs, including asymmetric DM~\cite{kaplan2009asymmetric}, hidden sector DM~\cite{foot2015dissipative}, and mirror DM~\cite{ignatiev2003mirror}. Direct detection experiments are also able to probe sub-GeV DM models through nuclear bremsstrahlung and Migdal effect signals~\cite{kouvaris2017probing,ibe2018migdal,dolan2018directly,akerib2018resultss}. In addition, forthcoming experiments will be sensitive to various astrophysical neutrino fluxes, most notably through the coherent nuclear scattering of solar, atmospheric and supernova neutrinos. This elusive elastic scattering process, which also generates very low energy recoils, has been recently observed for the first time with a pulsed neutrino beam ~\cite{akimov2017observation}. All of these interactions create steeply-falling energy spectra in direct detection detectors and any improvement in energy threshold, however modest, can bring about significant gains in sensitivity.\let\thefootnote\relax\footnotetext{$^*$ Corresponding authors: nellie.marangou15@imperial.ac.uk, \qquad  \linebreak tomaszbi@slac.stanford.edu}

The Large Underground Xenon (LUX) experiment completed two WIMP-search runs between 2013 and 2016, utilizing 250~kg of active liquid xenon (LXe) in a dual-phase xenon Time Projection Chamber (TPC)~\cite{Akerib2016,akerib2017results}. Its LXe target was surrounded by high-reflectance panels made from polytetrafluoroethylene (PTFE), and viewed by two arrays each with 61 photomultiplier tubes (PMTs, 2-inch Hamamatsu R8778). Energy depositions generated two distinct responses: a prompt scintillation signal (S1) and a delayed ionization response, the latter detected via electroluminescence of the drifted charge in a thin layer of vapor above the liquid (S2). The presence of the two distinct signals per interaction was particularly important for i) discriminating nuclear recoil (NR) signals from the prominent backgrounds, which consist mostly of electron recoils (ER), and ii) recovering the 3D position of the interactions, hence allowing self-shielding of an inner `fiducial volume' from external radioactivity backgrounds~\cite{LUXInst}. The gain factor for the prompt scintillation response, $g_1$, consists of the product of the detector light collection efficiency and the PMT quantum efficiency. The delayed ionization gain factor, $g_2$, is defined as the product of the average single electron response size and the electron extraction efficiency (from liquid to gas). Detailed descriptions of the experimental hardware and analysis techniques can be found in Refs.~\cite{LUXInst,akerib2018calibration} and others cited therein.

Although both the S1 and the S2 channels could sense very low recoil energies, the ionization channel was more sensitive and, in the standard (S1+S2) analysis, the S1 signal determined the energy threshold to a large degree. LUX has achieved thresholds of 1.2~keV for ER~\cite{akerib2016tritium} and 3.3~keV for NR~\cite{Akerib2016} (both quoted at 50\% efficiency) for a 2-fold coincidence requirement on S1 (i.e.,~a valid S1 must include detection of one or more photoelectrons in at least two PMTs)~\cite{akerib2018calibration}. The 2-fold coincidence eliminated a major source of background whereby PMT dark counts (DC) recorded up to a few hundred microseconds before an isolated S2 pulse faked a valid `golden' event possessing one S1 and one S2.

Traditionally, S1 and S2 pulse areas have been expressed as a total number of photoelectrons (phe) emitted by the PMT photocathodes, using as reference the mean responses of the full signal chain to the emission of a single photoelectron (SPE). Recently, it was found that at vacuum ultraviolet (VUV) wavelengths double photoelectron emission (DPE) --- whereby two `photoelectrons' are detected in response to a single incident photon --- can take place with sizeable probability~\cite{Faham2015}. For the detection of liquid xenon scintillation (mean wavelength 175~nm~\cite{xenon}) in cryogenic conditions this DPE probability is $\approx$17\% for the LUX PMTs, and as much as $\approx$23\% for the Hamamatsu R11410 model used in forthcoming experiments such as LUX-ZEPLIN (LZ)~\cite{paredes2018response}. In general, this probability varies significantly with PMT model and, in particular, with photocathode technology. The LUX pulse-parametrization algorithms were corrected to account for this effect, with the size of S1 and S2 pulses expressed in ‘photons detected’ (phd), instead of the traditional phe unit obtained by pulse charge integration. For larger pulses this involved an average correction factor per PMT, while for smaller (S1) pulses an algorithm for `spike' counting was introduced, where each spike was an excursion of the waveform above some (low) threshold. These corrections brought about significant improvements in both the linearity and the energy resolution of the detector~\cite{Akerib2016,akerib2016tritium}. 

\begin{figure}
\centering
\includegraphics[width=1
\linewidth]{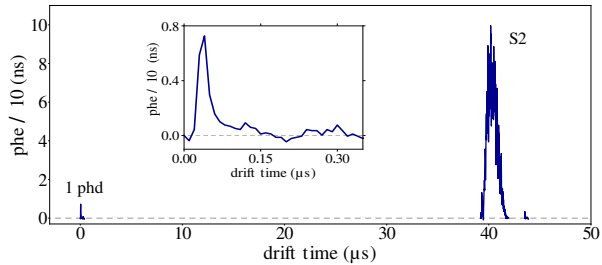}
\caption{A single S1-photon event from the Dec.~2013 LUX tritium dataset. The S2 pulse ($\sim$35 extracted electrons) was preceded by a pulse integrating to 2.4~phe --- likely due to DPE emission in response to a real S1 detected photon (the relevant PMT waveform is shown in inset). In the standard S1+S2 analysis this would have been classified as an `S2-only' event, as the S1 candidate has failed the 2-fold requirement~\cite{akerib2016tritium}.}
\label{fig:example_event}
\end{figure}

In this work we extended the standard `2-fold' LUX analyses by including single photon pulses (1~phd) in those instances where one VUV photon had produced two photoelectrons (2~phe) through DPE. This allowed lower energies to be detected by recovering a number of events that were previously discarded by the 2-fold requirement. PMT dark counts consist almost entirely of 1~phe signals and hence did not pose a significant background. This method retains $z$-position (depth) information and ER-NR discrimination for these events, and hence can provide a bridge between standard analyses with both S1 and S2 signals and the more challenging `S2-only' analyses~\cite{agnes2018low,angle2011search}, that can suffer from higher background levels (e.g.~from radioactivity or spurious electron emission from electrode grids~\cite{tomas2018study}). An example event that fell below the 2-fold requirement but in which the S2 was preceded by a single detected photon of large pulse area (indicative of DPE emission) is shown in Figure~\ref{fig:example_event}. Note the significant S2 pulse size for this small ($\sim$1~keV) ER signal; although NR signals generate proportionally smaller S2s, they still represent several emitted electrons that are detectable by electroluminescence.

For this to be a useful analysis we initially confirmed the efficient detection of low energy interactions generating S1 pulses consisting of single photons that have produced 2-phe signals. Additionally, we established that no major new backgrounds exist in this regime that could jeopardise a rare event search. For example, these could arise from even faint sources of VUV photons unconfined to S1 or S2 pulses (e.g. associated with electron transport in the liquid or electrical breakdown in liquid or gas phases) or from PMT-related spurious signals.

In this article we set out a full single-photon analysis of the LUX 2013 WIMP search dataset to arrive at a WIMP-nucleon scattering cross section limit using exclusively the population of events that have produced two photoelectron signals in a single PMT. This extends the standard analysis published previously in Ref.~\cite{Akerib2016}. We begin by describing the calibrations of single and double photoelectron responses of individual PMTs and their dark count rates (Section~II). In Section~III we validate this technique using a high-statistics ER calibration dataset obtained with tritiated methane (CH$_3$T) dispersed uniformly throughout the detector, extending the original analysis of this dataset~\cite{akerib2016tritium} to lower energy signals. This allows optimization of the S1 event selection and data corrections, and assessment of the DPE event detection efficiency. In Section~IV we describe our search for low-energy NR interactions in the LUX 2013 WIMP search dataset, producing spin-independent WIMP-nucleon cross section limits using DPE events alone. In Section~V we discuss our results and review the potential of this new analysis, in particular in the context of LZ \cite{LZTDR} (a separate publication will discuss this in detail). Next-generation LXe-TPCs will achieve extremely low backgrounds over multi-tonne masses and any threshold improvements can enable new physics.

\section{Photomultiplier calibration \label{sec:PMTs}}

We began by calibrating the SPE response, the DPE probability for LXe scintillation photons, and the dark count rates per PMT. We also identified and excluded noisy channels from our analysis.  We used data from a tritium calibration (Dec.~2013 injection), with a low event rate (peaking at $\sim$0.1~cts/s), allowing PMT dark counts to be studied. This dataset also provided small VUV pulses from particle interactions to calibrate the DPE probabilities.

\begin{figure}[h]
\centering
\includegraphics[width=1
\linewidth]{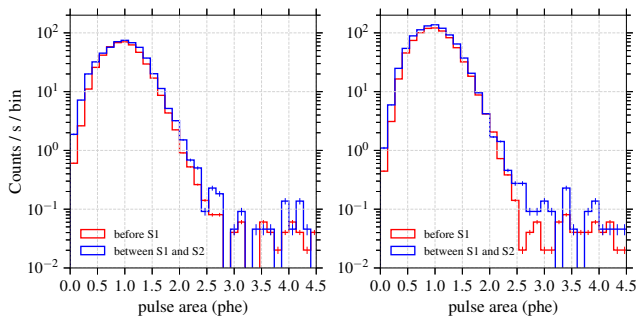}
\caption{Pulse area distributions (in phe) of SPE-like pulses summed across the top (left) and bottom (right) arrays. These were identified in the quiet waveforms preceding tritium S1 pulses (in red), or between the tritium S1 and S2 pulses (in blue). These distributions do not include contributions from 10 channels that were removed for this analysis.}
\label{fig:spe_response}
\end{figure}

Nominally, an SPE generated a pulse with 4 mV amplitude for the average PMT gain of 4$\times10^6$. The channel by channel SPE calibration was originally performed using the internal LED calibration system. The 400 nm light pulses used do not produce double photoelectron emission, causing a bias that was corrected at the analysis stage. In this work, the SPE responses were determined in two ways (both directly utilizing the tritium calibration dataset): firstly, by using a population of waveforms consisting mostly of SPE pulses; secondly, by examining the full single VUV photon response (including both SPE and DPE components). In the first case, pulses classified as SPE-like were used, and found either preceding an S1 or identified between the S1 and the S2 pulses and hence consisting mostly of PMT dark counts. A distribution is shown in Figure~\ref{fig:spe_response}, where all channels in the top and bottom PMT arrays have been summed for the purpose of display. The distributions for individual channels were fitted with a Gaussian model using the likelihood method to obtain the mean and width of the SPE response ($\mu_{DC}$,$\sigma_{DC}$), as well as the DC rate for each PMT. The fit was restricted to start at 0.3~phe to allow for the digitization threshold (information is saved only if the amplified signal exceeds 1.5~mV~\cite{akerib2018calibration}).

Ten PMT channels with outlier fit parameters, goodness-of-fit estimators ($\chi^2$), or abnormally high dark count rates --- often found to be detecting spurious light --- were not used for this analysis. The average dark count rates for the `good' PMTs, measured both using the tritium calibration and the WIMP search dataset, are summarized in Table~\ref{table:DC}. Several interesting observations result from these data, which are also apparent in Figure~\ref{fig:spe_response}. The DC rate recorded at the bottom array is significantly higher than that at the top array; and both are much lower than the $\sim$40~cts/s/PMT observed at low temperature in tests on the surface. These effects are also observed in other experiments~\cite{aprile2013x}. The latter effect is presumably due to the low radiation environment in LUX (e.g.~reducing Cherenkov emission in the PMT windows). The difference between the two arrays is more interesting; since the bottom array had a higher photon collection efficiency for interactions in the liquid, we believe that the rate measured there consists partly (or even mostly) of detected light, and is not attributed solely to thermionic emission from the photocathodes. This light did not undergo double photoelectron emission and hence was not of VUV wavelengths. A likely explanation is that these photons are due to PTFE fluorescence at optical wavelenghts, which is induced when VUV photons are absorbed in the PTFE (see~\cite{PTFE,PTFE2}). Other manifestations of this effect have previously been observed in LUX. This ability to distinguish between VUV and visible photon fluxes is a useful application of the DPE effect. Finally, there is a small difference between the rates measured in the two time windows (before S1 and between S1 and S2 pulses), both in measurements using the tritium dataset and during the WIMP search, which is unlikely to be statistical in nature. This appears at a small time window following the S1 and is more pronounced after larger S1 pulses, indicating it is likely due to S1-induced PTFE fluorescence. In the remainder we continue to refer to these SPE signals as `dark counts', but bearing in mind that most are actually photon-induced.

\begin{table}[h]
\caption{Average dark count rates for the PMTs used in this analysis as measured both via the tritium calibration and the 2013 WIMP search dataset. These rates were measured using populations of golden events, searching before the S1 pulse ($<$S1 window) and between the S1 and S2 pulses (S1--S2).}
\label{table:DC}
\begin{tabular}{c | l c c}
\hline
Window & Array & Tritium (cts/s) &  WS data (cts/s) \\
\hline
$<$S1 &  Top & \(8.2 \pm 0.4\)  & \(8.0 \pm 0.7\) \\
      &  Bottom & \(14.7 \pm 0.5\)  & \(14.0 \pm 0.9\) \\
\hline
S1--S2 &  Top & \(9.4 \pm 0.4\)  & \(9.6 \pm 3.1\) \\
       &  Bottom & \(16.8 \pm 0.6\)  & \(17.2 \pm 1.7\) \\
\hline
\end{tabular}
\end{table}

In general, there is reasonable agreement between the distributions before and after S1, and no significant excess is observed near two photoelectrons, which could indicate DPE emission from single photons. This confirms that no significant VUV photon sources exist outside of the main xenon luminescence mechanisms associated with S1 and S2 photon production, and hence no major backgrounds are expected that could make such an analysis non-viable --- this is one of the main results from our study. Nonetheless, the SPE distributions in Figure~\ref{fig:spe_response} do include a modest tail to large pulse areas, which we attribute to the way in which the LUX pulse classifier identifies SPE-type pulses, which is a loose requirement: those must be S1-like in shape and recorded in a single PMT within a 100~ns window, regardless of size~\cite{akerib2018calibration}. They would allow, for example, light pulses emitted within the PMTs themselves~\cite{amaudruz2018situ} and Cherenkov signals in the PMT windows --- in underground experiments the latter may be due to Compton electrons generated by background gamma-rays~\cite{araujo2012radioactivity}.

\begin{figure}
\centering
\includegraphics[width=1
\linewidth]{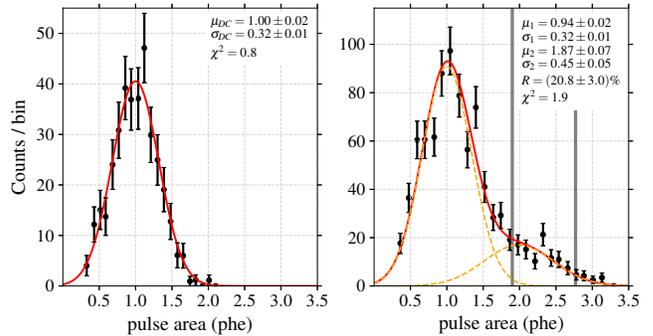}
\caption{Pulse area distributions (in phe) for the single photoelectron response (left) and the single VUV photon response (right) of an example PMT, along with the fit parameters. These include the mean and standard deviation obtained from the dark count population ($\mu_{DC}$,$\sigma_{DC}$), and the mean and standard deviation of both the SPE ($\mu_1$,$\sigma_1$) and DPE responses ($\mu_2$,$\sigma_2$) obtained from the single scintillation photon response. The DPE probability for this PMT is $R=(20.8\pm3.0$)\%. Grey vertical lines, in the right panel plot, indicate the signal region for this channel.}
\label{fig:fits}
\end{figure}

As mentioned previously, the full single VUV photon response was also measured, using VUV scintillation light from tritium interactions. Tritium $\beta^-$ decays generating very small S1 signals within appropriate small sub-volumes within the TPC were selected such that the mean expectation per channel was both small and constant ($\sim$0.02~phd), for the chosen range of S1s and sub-volumes. This expectation, and hence the probability for contamination from 2-phd signals, was calculated from the frequency of zero-hit events assuming Poisson statistics. An example of the single photoelectron and single photon responses (including both SPE and DPE components) for a typical PMT, fitted with Gaussian and double-Gaussian models, respectively, is shown in Figure~\ref{fig:fits}. We note that both the mean ($\mu_{DC}$,$\mu_{1}$) and the resolution ($\sigma_{DC}$,$\sigma_{1}$) of the two SPE responses are somewhat different, and this was observed for other channels. This effect is primarily attributed to the fact that the response measured with photons includes direct hits to the first dynode, biasing the mean of the distribution to lower values~\cite{Faham2015}. Additionally, in LUX a maximum of 10~pulses per event were parameterized in the standard data reduction, and small-area DCs may be lost; this effect increases the mean of the dark count SPE distribution by $\sim$2.5\%. In any case, this inefficiency for tagging SPE-like pulses has no effect on our DPE analysis ($<1\%$), as for the low-energy events of interest the 10 pulse limit is rarely reached. In addition, DPE pulses are larger than SPE pulses and hence more unlikely to be missed, as the pulse finder prioritizes larger pulses.

After correcting for a small ($\sim$2\%) contamination from 2-phd events, the mean DPE fraction was found to be $0.169 \pm 0.015_{sys} \pm 0.005_{stat}$, consistent with measurements from other LUX analyses. The statistical error presented here was propagated from the error on the fit value of the DPE fraction for each PMT (which were uncorrelated), while the systematic error was calculated by studying how the DPE fraction varied with S1 cut definition. The DPE fraction was found to vary significantly between different PMTs, confirming the need for the independent channel by channel calibration, with the sample standard deviation found to be 0.04.

\begin{figure}[h]
\centering
\includegraphics[width=1
\linewidth]{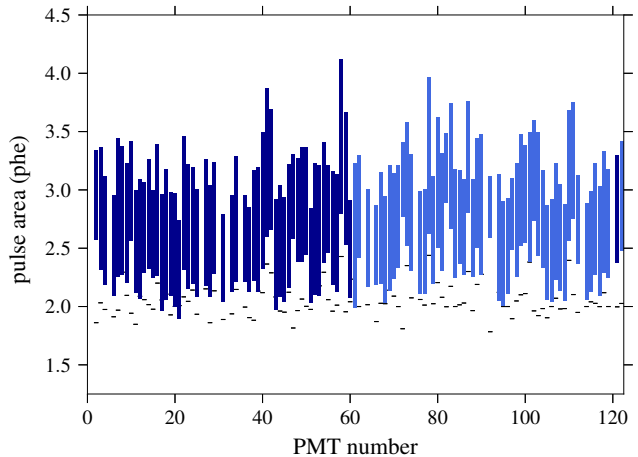}
\caption{DPE signal region for each active channel. Blue bars illustrate the [\(\mu_1 + 3 \sigma_1\), \(\mu_2 + 2 \sigma_2 \)] region for each PMT, while missing bars indicate where channels were not used for this analysis. The dark blue bars indicate channels located in the top PMT array, while the lighter blue bars show bottom array channels. Black lines show the mean of the DPE response ($\mu_2$) for each PMT.}
\label{fig:region}
\end{figure}

Given that the LUX PMTs did not fully resolve 1~phe from 2~phe signals, a cut must be applied that accepts DPE signals while rejecting dark count signals (we favoured the transparency of this simple cut-and-count method over more sophisticated procedures in this analysis). Figure~\ref{fig:fits} suggests that the optimal cut would be located near 2.0~phe. A signal-to-background ratio optimisation designed to limit the total number of background events to  $\mathcal{O}$(1) (given the DC and S2-only rates in the full 2013 WIMP search dataset) yielded the following acceptance region: $[\mu_1 + 3 \sigma_1,\mu_2 + 2 \sigma_2 ]$, where $\mu_i$ and $\sigma_i$ are the means and widths of the two Gaussians measured from the fits to the VUV dataset. Figure~\ref{fig:fits} (right) indicates this region for the PMT in question. The signal regions for all channels are shown in Figure~\ref{fig:region}. 

The rate of pulses classified as single photons that fall within the defined DPE signal region, over the entire analyzed array, was found to be (\(2.0 \pm 0.2\))~cts/s. This counting observation is in agreement with the expected rate due to dark count leakage into the signal region, that was calculated to be (\(2.2 \pm 0.2\))~cts/s, using the model fits. Using the counting observation, which could include a small contribution from random VUV single photon sources, ensures we assume all observed leakage rate in background expectation calculations and hence all possible sources are accounted for.

The mean acceptance for the single detected photon population recovered after the DPE cut for all good PMTs was found to be $0.055 \pm 0.005_{sys} \pm 0.002_{stat}$. This acceptance was weighted to take into account both the quantum efficiency of each PMT, the different coupling of the top and bottom arrays to scintillation light and the position of each PMT on the arrays (measured using $^{83\textrm {m}}$Kr calibration). As mentioned previously, 10 PMTs were turned off for this analysis, which resulted in a 6.0\% reduction in $g_1$ and a commensurate reduction in single photon event acceptance. Hence, with this applied, the DPE cut acceptance was estimated to be \(0.052 \pm 0.005 \). This acceptance translates directly to the analysis efficiency of the DPE cut, and hence represents the fraction of events with an S1 consisting of a single detected VUV photon that are recovered after its application.

\section{Extending the tritium ER calibration\label{sec:ERCalibration}}

Following the definition of the single-photon signal region, we proceeded to perform a single S1 photon calibration study using the 2013 tritiated methane (CH$_3$T) dataset. Tritium decays constitute an excellent calibration source for the new analysis, providing a large population of spatially uniform low-energy events; moreover, this beta spectrum is known with high precision. Significantly, the S1 event selection at 1~phd is identical for ER and NR datasets, and so it is directly applicable to the WIMP search analysis presented in the next section.

In the original analysis of the Dec.~2013 calibration dataset~\cite{akerib2016tritium}, $\sim$170,000 tritium decays were recorded in the chosen fiducial volume (radius less than 20~cm, drift time between 38 and 305~$\mu$s). For the new analysis, events with a single S2 pulse and no identified S1 pulse were initially selected and the standard S2 pulse quality and quiet time cuts were applied. Selected events were searched for a candidate single-photon pulse preceding the S2 and falling within the DPE acceptance region of the firing PMT --- such as the event already shown in Figure~1. For those fulfilling these requirements the 1~phd pulse was taken as the S1 signal, and all additional standard event selection cuts were applied \cite{Akerib2016,akerib2016tritium}. To benefit from the extension to lower energies, the (uncorrected) S2 threshold was lowered from 165~phd to 100~phd ($\sim$4 emitted electrons) and, consequently, a smaller fiducial volume (18~cm radial cut) was adopted for this analysis to avoid `wall' events being misreconstructed into the volume (this effect is a strong function of S2 size). The (x,y) position resolution for this S2 pulse area remains small ($\sigma \simeq 0.8$~cm~\cite{mercury}) and the trigger efficiency was $\approx$100\%~\cite{trigger}. Additional cuts removed pile-up events following large energy depositions in the detector and interactions occurring very near the liquid surface such that the S1 is essentially merged with the S2 pulse; these cuts had not been required in the standard analysis.

In total, $\sim$15,000 events passed the S2 quality cuts alone. These are largely ($\approx$75\%) due to genuine tritium decays with an S1 failing the 2-fold coincidence requirement, followed by the well-understood accidental coincidence backgrounds. Of those events, 247 were found with an S1 consistent with DPE emission on 1~phd within the appropriate drift time window. As mentioned previously, S2-only events may coincide with large-area DC pulses to produce a viable background topology. Using the rate of (\(2.0 \pm 0.2\))~cts/s for the latter, an expectation of \(10.8 \pm 1.1\) random coincidences was estimated for this tritium dataset.

\begin{figure}[h]
\centering
\includegraphics[width=0.95
\linewidth]{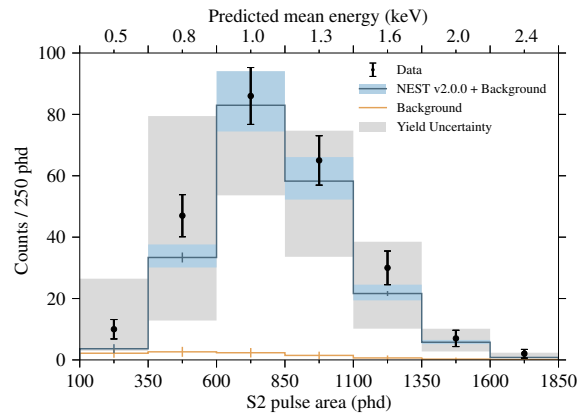}
\caption{S2 pulse area spectrum for tritium dataset events in which the S1 consisted of a single detected VUV photon with pulse area within the DPE signal region, along with NEST~v2.0.0 prediction added to the background expectation from the DC+S2-only coincidence events. The error bars on the NEST v2.0.0+Background line represent the uncertainty on the background expectation. The shaded blue regions represent the systematic uncertainty due to the $g_1$ and DPE cut acceptance error, while the shaded grey regions represent the yield uncertainties (produced by incorporating appropriate yield variations in the NEST model, shown in Figure~\ref{fig:yields}). The mean energy as predicted by NEST~v2.0.0 for the range of S2s in each bin is indicated on the top x-axis.}
\label{fig:s2}
\end{figure}

\begin{figure}[h]
\centering
\includegraphics[width=1\linewidth]{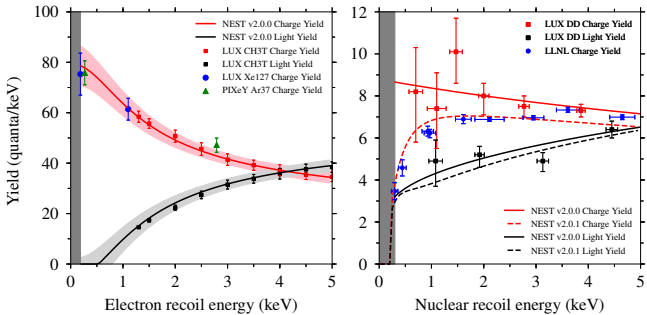}
\caption{The ER (left) and NR (right) ionization (red) and scintillation (grey) yields as a function of recoil energy given by NEST~v2.0.0. The bands around the ER yields indicate model uncertainties and correspond to those presented in Figure~\ref{fig:s2}, while red and black points indicate LUX measurements from the tritium (left)~\cite{akerib2016tritium} and D-D neutron (right)~\cite{akerib2016low} calibrations. Additional data are shown in blue and green for the ER ionization yield~\cite{boulton2017calibration,akerib2017ultralow}. Recent NR ionization yield measurements, performed at the Lawrence Livermore National Laboratory (LLNL), that have not yet been incorporated in the NEST v2.0.0 models but are used here to motivate the low energy threshold are presented in blue \cite{jingke}. The NEST v2.0.1 NR model yields, that became available after journal submission, incorporating the new data are shown using dotted lines. The thresholds adopted for the new analyses correspond to the rightmost part of the shaded regions.}
\label{fig:yields}
\end{figure}

The S2 pulse area spectrum for the single scintillation photon events passing all cuts is shown in Figure~\ref{fig:s2}. The S2 areas were corrected to account for spatial dependence. The Noble Element Simulation Technique (NEST) package v2.0.0~\cite{nest} was used to model the tritium S2 response for this selection and its prediction is plotted in the same figure, added to the (small) expected background distribution from upward-fluctuating DCs overlapping with S2-only events. Systematic and yield uncertainties are also shown. The former include errors on $g_1$ and the DPE cut acceptance added in quadrature, while the yield uncertainty includes ionization and scintillation yield variations within measurement errors at low energies --- see Figure~\ref{fig:yields}. The data are in good agreement with the NEST v2.0.0 ER model, and the total number of events predicted (inc.~background) is $208 \pm 21_{sys} \pm ^{25}_{11}$ ${}_{\!{yield}}$ (cf.~247 observed). The ER model predicts the energy of the single photon tritium events to lie between 0.3 and 2.6~keV, with a mean of 1.1 keV.

The ER and NR yields considered in NEST v2.0.0 are shown in Figure~\ref{fig:yields} along with key published measurements ~\cite{akerib2016tritium,boulton2017calibration,akerib2017ultralow,akerib2016low} that were used to obtain the  model fits. The ER model threshold presented here was placed at 0.186~keV, corresponding to the lowest energy ionization yield measurement~\cite{akerib2017ultralow}, and extends below the lowest energy at which the scintillation yield has been measured to date (1.3~keV~\cite{akerib2016tritium}). When extrapolating the model to our energy threshold, the scintillation signal was assumed to be anti-correlated with the charge signal as was observed at higher energies. The uncertainties presented as bands around the ER yields in Figure~\ref{fig:yields} indicate the variation observed when the free parameters of the model were allowed to vary by 1$\sigma$. The good agreement between data and model for the single photon ER calibration shown in Figure~\ref{fig:s2} suggests that the model extrapolations assumed in NEST, below the lowest light yield measurement, are reasonable for the energy range presented here.

\section{WIMP search\label{sec:Run3}}

The analysis of the tritium data confirmed that single scintillation photon events can be robustly selected to recover a population of very low energy interactions which had been previously discarded, and it suggested that the dominant random coincidence backgrounds can be calculated accurately. We hence applied this technique to search for light WIMP interactions in the 2013 WIMP search exposure which accumulated 95~live days~\cite{Akerib2016}.

We confirmed first that long-term PMT gain drift was not significant in this longer dataset (cf.~Fig.~20 in Ref.~\cite{akerib2018calibration}). From the PMT calibration work reported in Section~2 we were confident that no major sources of VUV photons or similar pulses were likely to be present, once the 10 noisy channels were excluded.

LUX has achieved a very low background rate especially at low energies~\cite{akerib2015radiogenic,akerib2018calibration}. Wall events were suppressed in the smaller fiducial mass of 118~kg (this fiducial definition is identical to that used in the first analysis of this dataset~\cite{akerib2014first}). The prominent background in standard analyses was instead due to ER interactions leaking into the NR region. Such ER backgrounds were small in the DPE  analysis due to the low DPE acceptance, and the same applies to NR backgrounds.
Table~\ref{table:backgrounds} summarizes the background expectations calculated for the leading sources along with the random coincidence background for the single photon 2013 WIMP search. The total prediction for 1~phd events was $5.1\pm0.4$. As the largest background is due to accidental coincidences between upward-fluctuating dark counts and S2-only events and the S2-only background spectrum is not flat, most of the background interactions are expected at S2 areas well above the NR signal region for light WIMPs.

\begin{table*}[t]
\caption{Expected backgrounds and observed counts for various ranges in S2 space for the single-photon WIMP-search analysis of the 2013 LUX WIMP search exposure (95~live days, 118~kg fiducial mass). Entries are related to an ER band calculated for a flat spectrum and defined at the 10$^\mathrm{th}$ and 90$^\mathrm{th}$ percentiles. S2 pulse sizes are given for spatially-corrected variables; the 100-phd uncorrected S2 threshold corresponds to $\simeq$120~phd.}
\label{table:backgrounds}
\begin{center}
\begin{tabular}{l c | c c c c c | c}
\hline
S2 region & S2 size (phd) & Coincidences   & ER         & NR         & Wall events  & Total  & Observed  \\ \hline
\(>\) 90\(\%\) ER & 1,110--5,000 & \(2.0 \pm 0.2\) & \(0.02 \pm 0.01\) & \(<0.01\) & \(<0.01\) & 
\(2.0 \pm 0.2\) & 3 \\  
10\(\%\) -- 90\(\%\) ER  & 515--1,110 & \(0.9 \pm 0.1\) & \(0.4 \pm 0.2\) & \(<0.01\) & \(<0.01\) & \(1.3 \pm 0.2\) & 1 \\  
\(<\) 10\% ER & $\simeq$120--515 & \(1.7 \pm 0.3\) & \(0.02 \pm 0.01\) & \(0.01 \pm 0.01\) & \(0.05 \pm 0.02\) & \(1.8 \pm 0.3\) & 2 \\  
\hline
Total & $\simeq$120--5,000 &  \(4.6 \pm 0.4\) & \(0.4 \pm 0.2\) & \(0.01 \pm 0.01\)& \(0.05 \pm 0.02\) & \(5.1 \pm 0.4\) & 6\\
\hline
\end{tabular}
\end{center}
\end{table*} 

\begin{figure}[h]
\centering
\includegraphics[width=1
\linewidth]{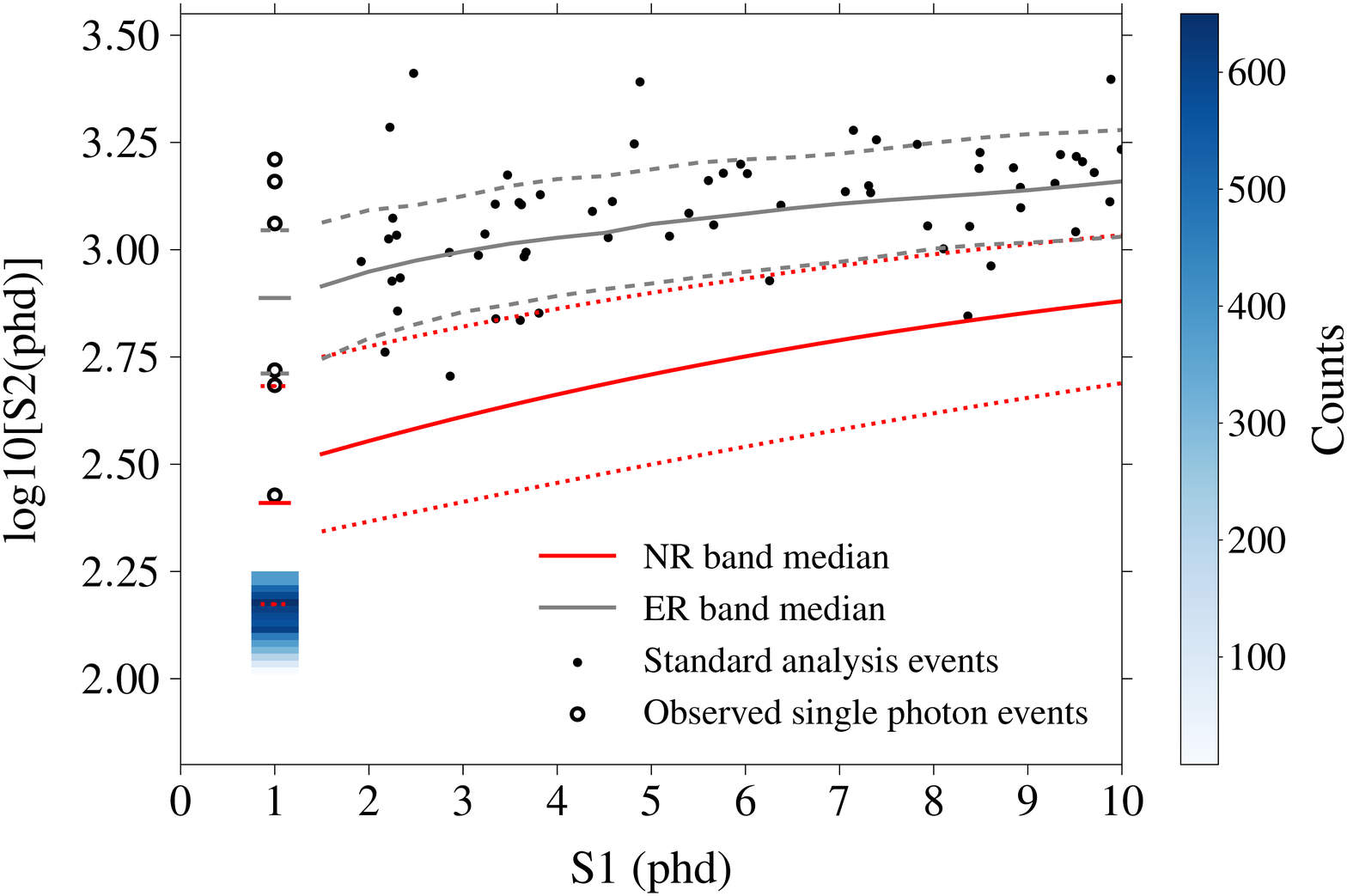}
\caption{Events observed in the 2013 LUX WIMP search exposure of 95~live days and 118~kg fiducial mass (11,210~kg$\cdot$days). Solid black markers represent events meeting the S1 2-fold coincidence requirement, while the 6~hollow markers indicate those with an S1 of 1~phd which are the focus of this analysis. Distribution contours for an ER beta spectrum (grey) and an example 50~GeV/c$^2$ WIMP signal (red) are indicated at the 50$^\mathrm{th}$ (solid), 10$^\mathrm{th}$ and 90$^\mathrm{th}$ (dashed) percentiles of S2 at given S1. These percentiles are shown separately at 1~phd, with the S2 threshold lowered to the $100$-phd value (uncorrected) adopted in this analysis. The color histogram illustrates the expected WIMP signal for a mass of 4~GeV/c$^2$ at 1~phd only, for an exposure of $10^6$~kg$\cdot$days and cross section of $10^{-40}$~$\text{cm}^2$; the integrated number of events for this model is $\sim$8,000.}
\label{fig:scatter}
\end{figure}

Figure~\ref{fig:scatter} shows the observation in (S1,\,S2) space both for interactions at $\geq$2-fold and for single-photon signals. A total of six single-photon events were observed. The numbers falling into each S2 region are consistent with the background expectations listed in Table~\ref{table:backgrounds}. Two events were observed below the ER band (expectation value was~$1.8\pm0.3$), i.e.~below an S2 pulse area which contains the full NR acceptance for WIMP signals.

In general, the NR signal region for light WIMPs lies below the NR band represented in the figure, which was derived for a 50~GeV/c$^2$ WIMP spectrum (which coincides approximately with the NR spectrum obtained with a D-D neutron generator). This is due to the fact that, at a particular S2 size, only events with an over-fluctuating S1 pulse fall above threshold. For the very low masses considered here this effect is even more extreme, with the predicted S2 range falling well below the NR band as defined for higher energy recoils. To determine the precise NR acceptance region for an S1 of 1~phd, WIMP signal models were simulated with NEST~v2.0.0 as a function of particle mass. The S2 acceptance regions extended between an S2 pulse area (uncorrected) of 100~phd and a maximum value that retains 95\% acceptance. A color histogram of the signal model for 4~GeV/c$^2$ WIMPs, with all analysis cuts and the acceptance region S2 cutoffs applied to it, is shown in Figure~\ref{fig:scatter}. The upper boundary of this region moves upward with increasing particle mass, and captures the first event for a 5.3~GeV/c$^2$ WIMP model at $\log_{10}(\text{S2})=2.4$.

The NR scintillation and ionization yields as a function of recoil energy as modeled in NEST v2.0.0 are presented in Figure~\ref{fig:yields}. The NR model cut-off was placed at 0.3 keV, which corresponds to the lowest-energy ionization yield data point in a newly published measurement~\cite{jingke}. Even though these NR measurements motivated in part this lower cut-off, they were not included in the NEST~v2.0.0 model development. A newer NEST model (NEST v2.0.1) that became available after this analysis had been completed is also shown in Figure~\ref{fig:yields}.  This was used to assess the effect of yield variations. We also present results with an energy threshold corresponding to the lowest light yield measurement (1.1~keV~\cite{akerib2016low}). A more detailed discussion of the NR NEST models and their effect on our result is presented in Section~5. 

\begin{figure}[h]
\centering
\includegraphics[width=0.95
\linewidth]{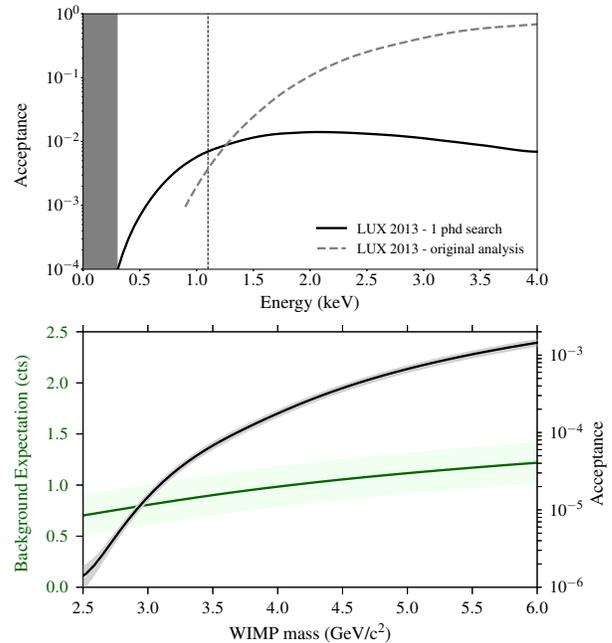}
\caption{\textit{Top}: Efficiencies for NR event detection as presented in the original analysis of the LUX 2013 data~\cite{Akerib2016} (dashed gray) and as estimated using NEST~v2.0.0 for the single-photon search presented here (black). These efficiencies include detection of single scatter events passing the S1 and S2 thresholds. In the original analysis the  S1 threshold required at least two PMTs detecting  photons and a minimum uncorrected S2 of 165 phd. The single photon analysis includes events with a single S1 photon producing double-photoelectron emission and an uncorrected S2 threshold of 100 phd.
\textit{Bottom}: Overall WIMP signal acceptance for the single photon analysis simulated with NEST~v2.0.0 after all cuts (black), along with background expectations (green) for the 2013 WIMP search exposure calculated for the appropriate acceptance region for each mass. Shaded green and black regions represent corresponding uncertainties.}
\label{fig:acceptance}
\end{figure}

Figure~\ref{fig:acceptance} shows the overall signal acceptance and background expectation in the appropriate S2 region as a function of WIMP mass, along with their uncertainties. The background expectation falls gently below 1~event at $\sim$4~GeV/c$^2$, while the acceptance decreases quickly due to a number of reasons. These include the S2 threshold of 100~phd (4 emitted electrons, corresponding to 8 electrons drifting in the liquid) and the hard signal cut-off which is applied at 0.3~keV in simulations. Finally, the lightest WIMPs increasingly produce genuine S2-only events at the lowest masses. The uncertainty on the signal acceptance includes both the uncertainty on $g_1$ and the DPE cut acceptance error. Yield uncertainties are not included in Figure~\ref{fig:acceptance} but the effect of yield variation is presented through the use of different yield models and discussed in Section~5. The uncertainty on the background expectation is determined by the position corrections on the S2 spectrum, which affects the number of events falling within the S2 acceptance region.

The statistical analysis technique described in Ref.~\cite{rolke2005limits} was used to set 90\% C.L. upper limits on the number of signal counts at each mass, using Gaussian uncertainties on the background expectation and the signal acceptance, as indicated in Figure~\ref{fig:acceptance}. These limits were capped at 2.3~counts below 5.3~GeV/c$^2$ since no counts were observed in the signal region for lower mass models, and this prevents our result for surpassing that of a background-free experiment. 

\begin{figure}[h]
\centering
\includegraphics[width=1
\linewidth]{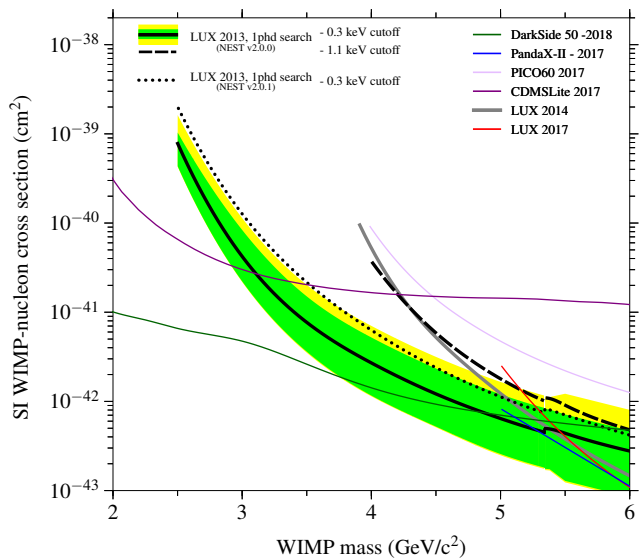}
\caption{90\% C.L.~upper limits on the spin-independent WIMP-nucleon cross section obtained using the single-photon population producing double-photoelectron emission in the LUX 2013 WIMP search. The observed limit with a 0.3~keV NR energy cut-off is shown in solid black, with 1$\sigma$ and 2$\sigma$ bands of background-only trials shown in green and yellow. The dashed black line is derived from the same analysis but with a model cut-off at 1.1~keV. Both of these results correspond to the NEST v.2.0.0 model shown in Figure~\ref{fig:yields}. The upper limit using a 0.3~keV NR energy cut-off with the newer NEST v.2.0.1 model is shown using a dotted black line. Also shown are the previous results from the LUX 2013 search~\cite{Akerib2016} (gray), the LUX complete exposure result~\cite{akerib2017results} (red), and from DarkSide-50~\cite{agnes2018low} (green), PandaX-II~\cite{cui2017dark} (blue), PICO60~\cite{amole2017dark} (lilac) and CDMSLite~\cite{agnese2015wimp} (purple).}
\label{fig:limit}
\end{figure}

From these results we calculated the corresponding upper limits on the spin-independent elastic scattering WIMP-nucleon cross section, assuming the same astrophysical parameters considered in~\cite{Akerib2016}; these limits are plotted in Figure~\ref{fig:limit}. The step seen at 5.3~GeV/c$^2$ marks the WIMP mass beyond which the signal region includes the first observed event. The expected sensitivity was calculated from background-only Monte Carlo trials and the 1$\sigma$ and $2\sigma$ regions are also shown. The median sensitivity follows approximately along the observed result. 

In conclusion, the new result is fully consistent with the background-only hypothesis. This analysis utilized exclusively single-photon events and not those at 2-fold and above, producing competitive results below 5.5~GeV/c$^2$. There are, however, differences between this and the original analysis that contribute to this improvement, as discussed below.

\section{Discussion\label{sec:Discussion}}

Inevitably, the new analysis demands a good understanding of scintillation and ionization yields at very low energies, and in particular the light yields for both ER and NR rely on models reaching below the lowest measurements at present. The LUX collaboration and others have been working for over a decade to extend these measurements and improve their systematic uncertainty, and major progress has been made for liquid xenon --- and this is set to continue.

The analysis of electron recoil interactions from tritium decays suggests that these yields are well modelled by NEST down to sub-keV energies already. In addition, we have shown that a population of single-photon events undergoing double-photoelectron emission can be reliably selected to lower the detection threshold for ER interactions. The modest backgrounds mean that this technique can be used in searches for leptophilic DM and other rare interactions producing electronic recoils near threshold in liquid xenon detectors.

The nuclear recoil analysis is equally promising, but it is important to note that the improvement in WIMP sensitivity over the previous 2013 WIMP search result is due to several effects. The lowering of the S1 threshold allowed probing lower energy recoils, and hence it improved the sensitivity at lower masses. Clearly, the background suppression due to the low acceptance of single photon events enabled a decrease in S2 threshold. However, these effects are not apparent when enforcing the 1.1~keV cut-off adopted in previous analyses. This threshold was motivated by the lowest-energy light yield measurement for NR interactions~\cite{akerib2016low} shown in Figure~\ref{fig:yields}. The result of applying this higher threshold to the 1-phd analysis is also shown in Figure~\ref{fig:limit}. (It is impractical to assess the effect of lowering the energy cut-off in the standard analysis instead, as this employed a different NEST model and a more powerful statistical analysis, with selection cuts optimized for that analysis.)

The lower (0.3 keV) model cutoff for NR interactions was adopted here as this is approximately the energy at which the first ionization electron is expected from extrapolating to very low energies the power-law behaviour that is assumed in the Lindhard model \cite{lindhard1963integral,lindhard1963integral,sorensen2011nuclear}. Recent measurements~\cite{jingke} confirm electron release at 0.3 keV in liquid xenon but are consistent with a lower ionization yield than the one assumed in the NEST v2.0.0 model. To understand how such yield variations affect the WIMP sensitivity, we also present an upper limit result obtained using the newer NEST v2.0.1 NR model, shown using dashed lines in Figure~\ref{fig:yields} (which became available shortly after journal submission). The NEST v2.0.1 NR model incorporates data from the recent publication and is also consistent at the 1$\sigma$ level with previous measurements conducted with LUX (as shown in Figure~\ref{fig:yields}). While the NEST v2.0.0 model assumes total scintillation and ionization yield anti-correlation below the smallest light yield measurement, the new NEST v2.0.1 model allows the total number of quanta to decrease at lower energies.
Use of the NEST v2.0.1 model resulted in a slightly higher observed limit (shown as a dotted black line in Figure~\ref{fig:limit}) but within the 1-2$\sigma$ band of the NEST v2.0.0 upper limit.


It is worth considering whether a DPE cut is required in the first place, i.e.,~would a full analysis including all 1-phd pulses be equally sensitive, despite the much higher background? For the tritium dataset most such pulses are indeed sub-threshold S1 signals, and we have confirmed that the background-subtracted S2 spectrum is still in good agreement with the NEST prediction. Clearly, in this case both analyses may be useful, although the signal-to-background ratio is markedly better with the DPE cut ($\sim$20 versus $\sim$3, in this instance). On the other hand, for the rare event search the conclusion is more nuanced. For the masses explored in Figure~\ref{fig:limit}, accepting all 1-phd pulses would increase the efficiency by $\sim$20-fold, while the background expectation would increase by a factor of 300--500 over the numbers in Figure~\ref{fig:acceptance}. A full analysis of the LUX dataset, including all 1-phd S1 events, yields comparable 90\%~C.L.~upper limits to that including the DPE cut (with the analysis without a DPE cut found to be a factor of $\sim$3 less sensitive at 4~GeV/c$^2$), but the sensitivity is now dominated by the systematic uncertainty on the background expectation. This is largely determined by the population of S2-only events that are predicted to fall within the S2 signal region for each mass: in the absence of meaningful depth ($z$) information, correcting the S2 pulse (e.g.~for finite electron lifetime) cannot be done accurately event-by-event. In addition, for such an analysis the possibility of events presenting both signal and background (one or more dark counts coinciding with a real 1-phd S1) becomes non-negligible and needs to be addressed. Moreover, systematic uncertainties play a bigger role in a discovery situation, and the new analysis is better able to control these. Also in this case, exploring both analyses in parallel is advisable.


Finally, we discuss the potential of this technique for future xenon experiments such as LZ (a detailed study will be published separately) \cite{LZTDR}. The sensitivity improvement due to the lowering of the S1 threshold alone is relatively modest here as LUX is already operating at a very low 2-fold coincidence level, and it is limited by both the S1 and the S2 thresholds. LZ can benefit further from this analysis technique by reducing its S1 coincidence requirement from 3-fold to 2-fold. In this instance, either or both detected photons can undergo double photoelectron emission, enabling a more efficient recovery of 2-fold events. Furthermore, the electron extraction efficiency and mean single electron size are relatively low in LUX, and low-energy searches in LZ should be driven predominantly by the S1 threshold. Additionally, the Hamamatsu R11410 PMTs used in LZ are found to have higher DPE probabilities which would also aid in a more efficient analysis. Finally, there are prospects for lower coincidence backgrounds from spurious electron and photon emission from its grids as this has been the subject of significant research in recent times~\cite{tomas2018study}.

\section{Conclusion\label{sec:Conclusion}}

We presented a new data analysis technique to search for rare electron and nuclear recoil interactions at sub-keV energies in LXe-TPCs. This analysis is based on the efficient detection of single VUV photons that occasionally generate two photoelectrons in some photomultiplier tube models. Although the dual photoelectron response is a modest fraction of the total response to single VUV photons, there is essentially a very small dark count rate competing with such signals and low backgrounds can therefore be achieved --- this is a key conclusion of this study. For electron recoils we demonstrated the accurate reconstruction of a population of events where the S1 pulse consisted of a single detected photon recorded in the tritium calibration of the LUX experiment. We then applied a similar methodology to a search for low-energy nuclear recoils in the LUX 2013 WIMP search dataset, improving the spin-independent scattering cross section limits significantly between 2.5~GeV/c$^2$ and 5~GeV/c$^2$ WIMP mass compared with the previous analysis --- where a standard 2-fold threshold had been applied to the S1 signal.

Various groups around the world are pursuing the measurement of scintillation and ionisation yields for ER and NR interactions and to establish the energy required to release the first quantum of ionisation and of scintillation in liquid xenon. Therefore, this technique has the potential for low systematic uncertainty for both types of interactions. There are good prospects for applying this analysis to larger experiments such as LZ, where the improvement can be more significant owing to several factors. This could be an important enhancement in the search for very light WIMP interactions and the coherent nuclear scattering of solar neutrinos.

\section{Acknowledgements\label{sec:acknowledgements}}
We wish to thank Ibles Olcina and Alastair Currie for useful discussions. 

This work was partially supported by the U.S. Department of Energy (DOE) under award numbers DE-AC02-05CH11231,    DE-AC05-06OR23100, DE-AC52-07NA27344, DE-FG01-91ER40618, DE-FG02-08ER41549, DE-FG02-11ER41738, DE-FG02-91ER40674, DE-FG02-91ER40688, DE-FG02-95ER40917, DE-NA0000979, DE-SC0006605, DE-SC0010010,   DE-SC0015535, and  DE-SC0019066; the U.S. National Science Foundation under award numbers PHY-0750671, PHY-0801536, PHY-1003660, PHY-1004661, PHY-1102470,  PHY-1312561, PHY-1347449, PHY-1505868, and  PHY-1636738; the Research Corporation grant RA0350; the Center for Ultra-low Background Experiments in the Dakotas (CUBED); and the South Dakota School of Mines and Technology (SDSMT). LIP-Coimbra acknowledges funding from Funda\c{c}\~{a}o para a Ci\^{e}ncia e Tecnologia (FCT) through project-grant PTDC/FIS-PAR/28567/2017. Imperial College and Brown University thank the UK Royal Society for travel funds under the International Exchange Scheme (IE120804). The UK groups acknowledge institutional support from Imperial College London, University College London and Edinburgh University, and from the UKRI Science \& Technology Facilities Council for PhD studentships M126369B (NM) and T93036D (RT). The University of Edinburgh is a charitable body, registered in Scotland, with registration number SC005336.
 
This research was conducted using computational resources and services at the Center for Computation and Visualization, Brown University.

We gratefully acknowledge the logistical and technical support and the access to laboratory infrastructure provided to us by the Sanford Underground Research Facility (SURF) and its personnel at Lead, South Dakota. SURF was developed by the South Dakota Science and Technology authority, with an important philanthropic donation from T. Denny Sanford. Its operation is funded through Fermi National Accelerator Laboratory by the Department of Energy, Office of High Energy Physics.

\bibliographystyle{apsrev4-1}
\bibliography{bibfile}

\end{document}